\documentclass[a4paper,fleqn,usenatbib]{mnras}
\usepackage{lscape}
\usepackage{hyperref}

\usepackage[T1]{fontenc}
\usepackage{ae,aecompl}

\usepackage{graphicx}	
\usepackage{amsmath}	
\usepackage{amssymb}	

\title[The most metal-poor substellar object]{Primeval very  low-mass stars and brown dwarfs -- II. The most metal-poor substellar object}

\author[Z. H. Zhang et al.]{Z. H. Zhang,$^{1,2}$\thanks{E-mail:
zenghuazhang@hotmail.com}\thanks{
Based on observations collected at the European Organisation for Astronomical Research in the Southern Hemisphere under ESO programme 098.D-0222.}  D. Homeier,$^{3}$ D. J. Pinfield,$^{4}$  N. Lodieu,$^{1,2}$  H. R. A. Jones,$^{4}$
\newauthor 
F. Allard$^{5}$ and Ya. V. Pavlenko$^{6}$
\\
$^{1}$Instituto de Astrof{\'i}sica de Canarias, E-38205 La Laguna, Tenerife, Spain \\
$^{2}$Universidad de La Laguna, Dept. Astrof{\'i}sica, E-38206 La Laguna, Tenerife, Spain \\
$^{3}$Zentrum f{\"u}r Astronomie der Universit{\"a}t Heidelberg, Landessternwarte, K{\"o}nigstuhl 12, D-69117 Heidelberg, Germany  \\
$^{4}$Centre for Astrophysics Research, Science and Technology Research Institute, University of Hertfordshire, Hatfield AL10 9AB, UK \\
$^{5}$Univ Lyon, ENS de Lyon, Univ Lyon 1, CNRS, Centre de Recherche Astrophysique de Lyon, UMR5574, F-69007, Lyon, France \\
$^{6}$Main Astronomical Observatory, Academy of Sciences of the Ukraine, Golosiiv Woods, 03680 Kyiv-127, Ukraine 
}

\date{Accepted 2017 February 7. Received 2017 February 6; in original form 2016 November 27}

\begin{document}

\label{firstpage}
\pagerange{\pageref{firstpage}--\pageref{lastpage}}
\maketitle
\begin{abstract}
SDSS J010448.46+153501.8 has previously been classified as an sdM9.5 subdwarf. However, its very blue $J-K$ colour ($-0.15 \pm 0.17$) suggests a much lower metallicity compared to normal sdM9.5 subdwarfs. Here, we re-classify this object as a usdL1.5 subdwarf based on a new optical and near-infrared spectrum obtained with X-shooter on the Very Large Telescope. Spectral fitting with BT-Settl models leads to $T_{\rm eff}$ =  2450 $\pm$ 150 K, [Fe/H] = $-$2.4 $\pm$ 0.2 and log $g$ = 5.5 $\pm$ 0.25. We estimate a mass for SDSS J010448.46+153501.8 of 0.086 $\pm$ 0.0015 M$_{\sun}$ which is just below the hydrogen-burning minimum mass at [Fe/H] = $-$2.4 ($\sim$0.088 M$_{\sun}$) according to evolutionary models. Our analysis thus shows SDSS J0104+15 to be the most metal-poor and highest mass substellar object known to-date. We found that SDSS J010448.46+153501.8 is joined by another five known L subdwarfs (2MASS J05325346+8246465, 2MASS J06164006$-$6407194, SDSS J125637.16$-$022452.2, ULAS J151913.03$-$000030.0 and 2MASS J16262034+3925190) in a `halo brown dwarf transition zone' in the $T_{\rm eff}$--[Fe/H] plane, which represents a narrow mass range in which unsteady nuclear fusion occurs. This halo brown dwarf transition zone forms a `substellar subdwarf gap' for mid L to early T types.

\end{abstract}




\section{Introduction}
Theoretical studies have shown that primordial Pop III stars were predominantly very massive \citep[$M \ga$ 100 M$_{\sun}$;][]{brom02,yosh06}. However, \citet{chie01} and \citet{sies02} have reported a mechanism to form metal-free intermediate and low-mass stars ($M$ = 1--8 M$_{\sun}$), and more recently numerical simulations have demonstrated that metal-free stars with masses down to $\sim$0.1 M$_{\sun}$ can form due to recurrent/periodic gravitational instability \citep{clar11,grei11,basu12}. The initial mass function at 0.01--4 M$_{\sun}$ (including brown dwarfs and stars) is likely independent of metallicity within 0.01--3 Z$_{\sun}$, according to numerical simulations of  star formation from turbulent cloud fragmentation \citep{bate14}. 

Searches for very metal-poor \citep[VMP, $\rm {-3 < [Fe/H]} < -2$;][]{beer05} and Pop III stars have to-date generally focused on F- and G-type dwarfs, and G- and K-type turn-off stars, which are bright and can be studied fairly easily with high-resolution optical spectra (for metallicity determination). The majority of known VMP dwarf and giant stars have masses of 0.6--0.8 and 0.8--1.0 M$_{\sun}$, respectively. Very low mass stars (VLMS; $M  \approx$ 0.08--0.5 M$_{\sun}$) that are 4--10 mags fainter, have not previously been specifically targeted for VMP and Pop III stars in general. Although VLMS is the most numerous population, the number of known M-type VMP stars \citep{giz97,bur06,lep08,zha13,kir16,lod17} is significantly smaller than that of F- and G-type VMP stars \citep[e.g.][]{soub16}. Meanwhile, substellar object with [Fe/H] $\la$ --2.0 has not been reported in the literature to-date. 

The nuclear fusion in VLMS is dominated by the pp I chain reaction, which fuses hydrogen in the central part of VLMS, and the reaction efficiency is lower in stars with lower masses. Therefore, VMP VLMS reflecting the chemical composition of the gas from which they formed. They could provide crucial clues to the star formation history and the synthesis of chemical elements in the early Universe. M subdwarfs have masses in the range $\sim$ 0.09--0.5 M$_{\sun}$ and represent the majority of metal-deficient VLMS,  according to the mass function of the Galactic halo \citep[e.g. fig. 8 of][]{cha03}. L subdwarfs are expected to be a mixture of the least massive metal-deficient stars and brown dwarfs across the hydrogen-burning minimum mass \citep[HBMM; $\sim$ 0.08--0.087 M$_{\sun}$, depending on metallicity;][]{bara97,cha97}. The most metal-poor L subdwarfs are particularly interesting, because they represent low-mass stellar and substellar formation within an extremely low-metallicity environment.

There are currently 36 L subdwarfs reported in the literature (see table 4 in \citealt{zha17} and table 4 in \citealt{lod17}). L subdwarfs are classified into three metallicity subclasses, subdwarf (sdL), extreme subdwarf (esdL) and ultra subdwarf (usdL), based on optical and near-infrared (NIR) spectra \citep{zha17}, that extends and follows the nomenclature of subclasses of M subdwarfs \citep{lep07}. The metallicity ranges of  usdL, esdL, and sdL subclasses are: [Fe/H] $\lid$ --1.7,  --1.7 < [Fe/H] $\lid$ --1.0 and --1.0 < [Fe/H] $\lid$ --0.3, respectively. The five most metal-poor objects were re-classified as L ultra subdwarfs (usdLs), including 2MASS J16262034+3925190 \citep[2MASS J1626+39, usdL4;][]{bur04}, SSSPM J10130734--1356204 \citep[SSSPM J1013--13, usdL0;][]{sch04}, SDSS J125637.16$-$022452.2 \citep[SDSS J1256$-$02, usdL3;][]{siv09}, ULAS J135058.86+081506.8 \citep[usdL3;][]{lod10} and WISEA J213409.15+713236.1 \citep[usdL0.5;][]{kir16}. 
Using the most advanced ultracool model atmospheres \citep[e.g. BT-Settl;][]{alla14}, it is possible to constrain the metallicity ([Fe/H]) of VLMS at a precision of $\sim$0.2 dex, by fitting models to the full optical--NIR spectra with $\lambda/\Delta\lambda \ga 120$. 2MASS J1626+39, SSSPM J1013--13, and SDSS J1256$-$02 all have [Fe/H] = --1.8 $\pm$ 0.2 according to the BT-Settl model fits \citep{zha17}. 

SDSS J010448.46+153501.8 (SDSS J0104+15) was  selected from the Sloan Digital Sky Survey \citep[SDSS;][]{yor00} and the UKIRT Infrared Deep Sky Survey \citep[UKIDSS;][]{law07}. It was classified as sdM9.5 based on a low-resolution optical spectrum \citep{lod17}, according to an M subwarf classification scheme \citep{lep07}. However, metallicity consistency across the subclasses of this scheme has not been tested for the later M subtypes, and \citet{zha17} found that the late type sdMs (within the \citealt{lep07} scheme) are actually more metal-poor than early type sdMs. By comparing the $i-J$ and $J-K$ colours of SDSS J0104+15 to an expanded M and L subdwarf sample in \citet{zha17}, we found that SDSS J0104+15 could have a significantly lower metallicity than suggested by the sdM9.5 type. We therefore obtained a new high-quality optical to NIR spectrum of SDSS J0104+15 to re-address its metallicity and classification.

This is the second paper of a series under the title {\sl `Primeval very low-mass stars and brown dwarfs'}.  In the first paper of the series, we reported the discovery of six new L subdwarfs, defined a new classification scheme for L subdwarfs and derived the atmospheric properties of 22 late type M and L subdwarfs \citep{zha17}. 
The observations of SDSS J0104+15 are presented in Section \ref{sobs} of this paper. Section \ref{spro} presents constraints of characteristics of SDSS J0104+15, and discussions on the HBMM and the halo brown dwarf transition zone. Finally Section \ref{ssac} presents a discussion of our results.

\begin{figure}
\begin{center}
 \includegraphics[width=\columnwidth]{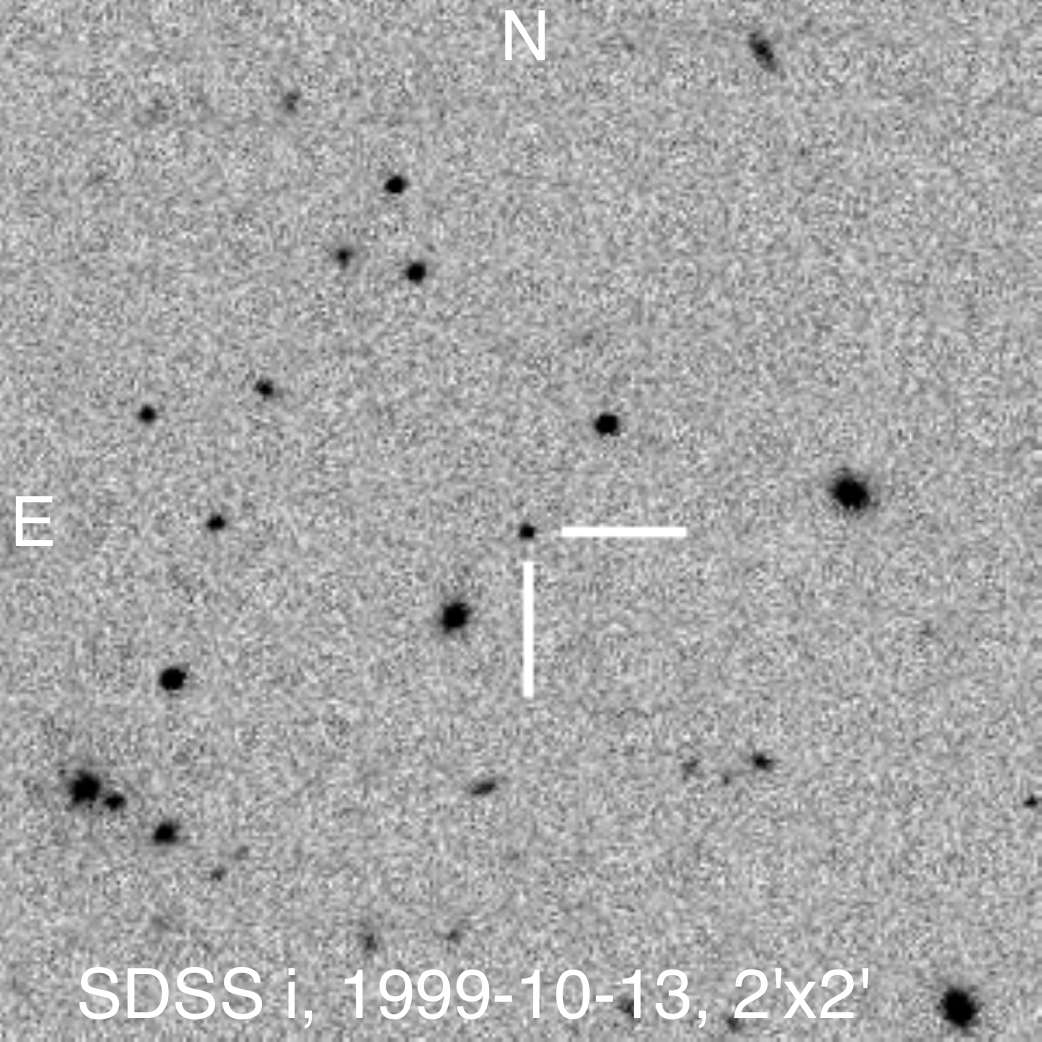}
\caption{SDSS $i$-band image of the field centred on SDSS J0104+15 (observation date 1999 October 13). The field is 2 arcmin on a side with north up and east to the left.}
\label{sdssi}
\end{center}
\end{figure}

\begin{figure*}
\begin{center}
   \includegraphics[angle=0,width=\textwidth]{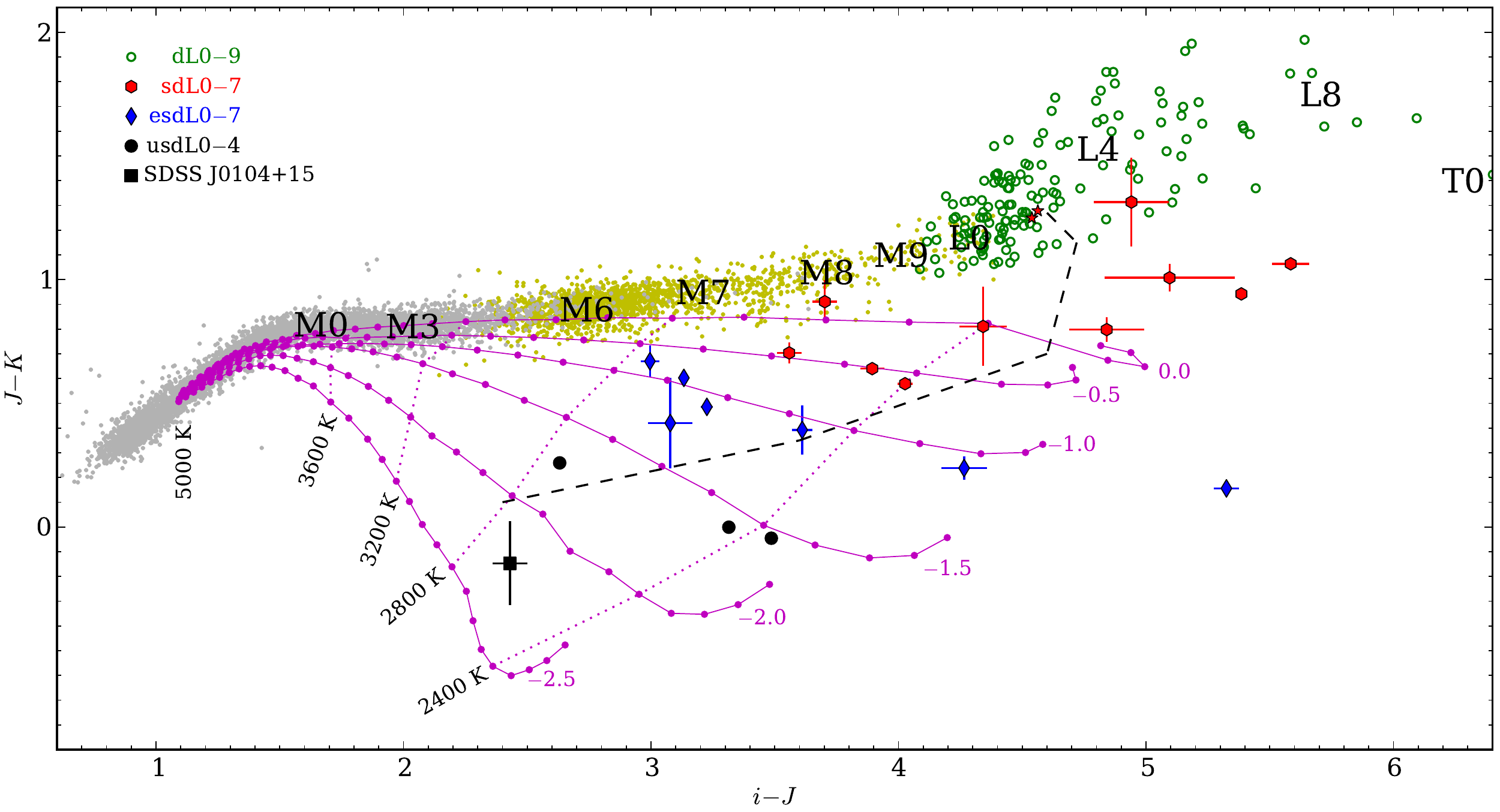}
\caption[]{The $i-J$ versus. $J-K$ colours of L subdwarfs compared to M and L dwarfs.  Red hexagon, blue diamonds and black circles are sdL, esdL and usdL subdwarfs classified by \citet{zha17}. The three usdLs (black circles from left to right) are SSSPM J1013$-$13 (usdL0), SDSS J1256$-$02 (usdL3) and 2MASS J1626+39 (usdL4). The black square is SDSS J0104+15. Some objects do not show error bars because these are smaller than the symbol size. Grey dots are 5000 point sources selected from a 10 deg$^2$ area of UKIDSS sky with $14<J<16$. Yellow dots are 1820 spectroscopically confirmed late type M dwarfs (for which mean spectral types are indicated) from \citet{wes08}. The BT-Settl model grids \citep{alla14} with log $g$ = 5.5 (magenta) are over plotted for comparison, with $T_{\rm eff}$ (2000--5000 K) and [Fe/H]  (from $-$2.5 to 0.0) indicated. Two five-pointed stars filled with red are the L1 SSSPM J0829$-$1309 and the L2.5 2MASS J0523$-$1403, which are likely least massive stars in the local field \citep{diet14}. They are not detected in SDSS and UKIDSS; therefore, PS1 $i$, and VHS $J$ and $K$ photometry are used. The difference between SDSS $i$ and PS1 $i$-band photometry of L dwarfs is $\sim \pm$0.05. The black dashed broken line indicates the roughly stellar--substellar boundary. Note this boundary is based on observed colours of least massive stars and brown dwarfs, not based on model predicted colours. } 
\label{ijk} 
\end{center}
\end{figure*}

\section{Observations}
\label{sobs}
\subsection{Photometry}
SDSS J0104+15 was first detected in the IR band by the Digitized Sky Survey II on 1992 September 25. It was also detected by the SDSS in the $r$, $i$ and $z$ bands on 1999 October 13, and by the UKIDSS Large Area Survey (ULAS) in the $Y$ and $J$ band on 2008 October 20, and in the $H$ and $K$ bands on 2007 November 25. It was detected by the {\sl Wide-field Infrared Survey Explorer} \citep[{\sl WISE};][]{wri10} in the $W1$ and $W2$ bands on 2010 July 13. It was observed by the Pan-STARRS1 \citep[PS1;][]{cham16} in the $i_{\rm P1}$, $z_{\rm P1}$, and $y_{\rm P1}$ bands with a mean epoch on 2012 December 27. Fig. \ref{sdssi} shows the SDSS $i$-band finder chart of SDSS J0104+15. It was selected as an ultracool subdwarf candidate by its red $i-J$ and blue $J-K$ colours, and was classified as an sdM9.5 subdwarf based on an optical spectrum ($\lambda/\Delta\lambda \approx 350$) obtained with the FOcal Reducer and low dispersion Spectrograph 2 \citep[FORS2;][]{app98} on the Very Large Telescope (VLT) on 2012 November 07 \citep{lod17}. 

Fig. \ref{ijk} shows the $i-J$ and $J-K$ colours of L subdwarfs compared to those of main sequence stars and brown dwarfs, with BT-Settl model colours \citep{alla14} over plotted. SDSS J0104+15 is located below and to the left of the three previously known usdL subdwarfs, indicating that SDSS J0104+15 could have a lower metallicity.  However, the low-resolution FORS2 optical spectrum is not good enough (in terms of wavelength coverage and resolution) for tight constraints of $T_{\rm eff}$, [Fe/H] and radial velocity (RV) of  SDSS J0104+15. 

\subsection{VLT spectroscopy} 
\begin{figure*}
\begin{center}	
 \includegraphics[width=\textwidth]{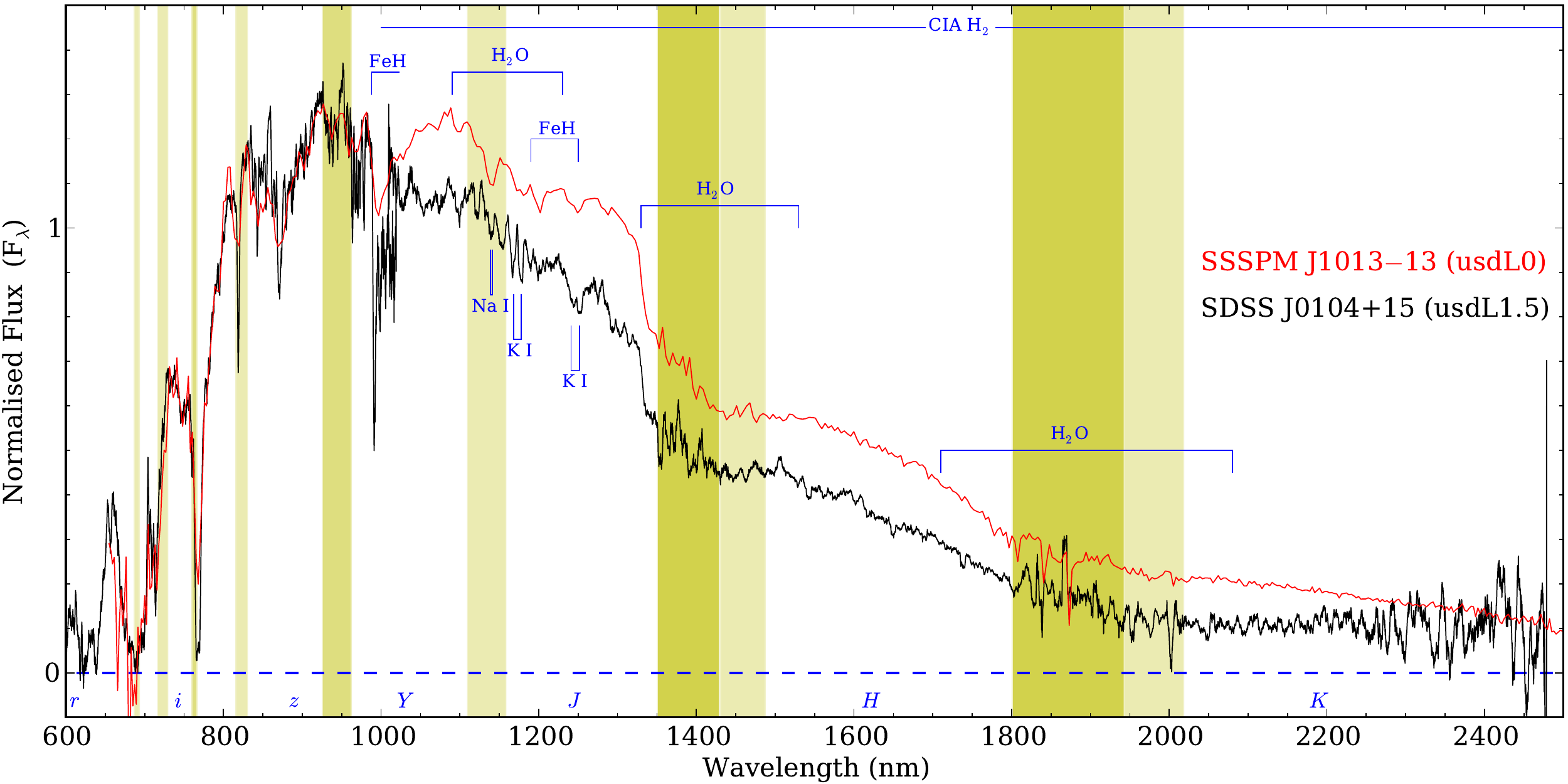}
\caption[]{The optical--NIR spectrum of SDSS J0104+15 compared to SSSPM J1013$-$13. The spectrum of SSSPM J1013$-$13 is from  \citet{bur04}.  Spectra are normalized near 800 nm. The spectrum of SDSS J0104+15 was smoothed by a boxcar function of 101 pixels to increase the signal to noise ratio. Telluric absorption regions are highlighted in yellow and have been corrected in our X-shooter spectrum. Lighter and thicker shaded bands indicate regions with weaker and stronger telluric effects.}
\label{spnir}
\end{center}
\end{figure*}

We obtained an optical to NIR spectrum of SDSS J0104+15 with X-shooter \citep{ver11} on the VLT on 2016 September 10 under excellent seeing conditions (0.43 arcsec as measured by differential image motion seeing monitor) and an average airmass of 1.7. The X-shooter spectrum was observed in an ABBA nodding mode with a 1.2 arcsec slit which provides a resolving power of 6700 in the VIS arm and 4000 in the NIR arm. The total integration time was 3480 s in the visible (VIS) and 3600 s in the NIR. A wavelength and flux calibrated 2D spectrum of SDSS J0104+15 was reduced with European Southern Observatory (ESO) Reflex \citep{freu13}. The 1D spectrum was extracted from the 2D spectrum with {\scriptsize IRAF}\footnote{IRAF is distributed by the National Optical Observatory, which is operated by the Association of Universities for Research in Astronomy, Inc., under contract with the National Science Foundation.} task {\scriptsize APSUM}. Telluric correction was achieved using the B9 star HD182719 which was observed a few minutes before SDSS J0104+15 at an airmass of 1.64. The spectrum of SDSS J0104+15 has signal-to-noise (SNR per pixel) of $\sim$29 at 800 and $\sim$10 at 1300 nm.  Spectra plotted in Fig. \ref{spnir} are smoothed by 101 pixels (boxcar smooth with {\scriptsize IRAF SPLOT}), which increased the SNR by a factor of 10 and reduced the resolving power to $\sim$ 600--400.

\begin{figure}
\begin{center}
 \includegraphics[width=\columnwidth]{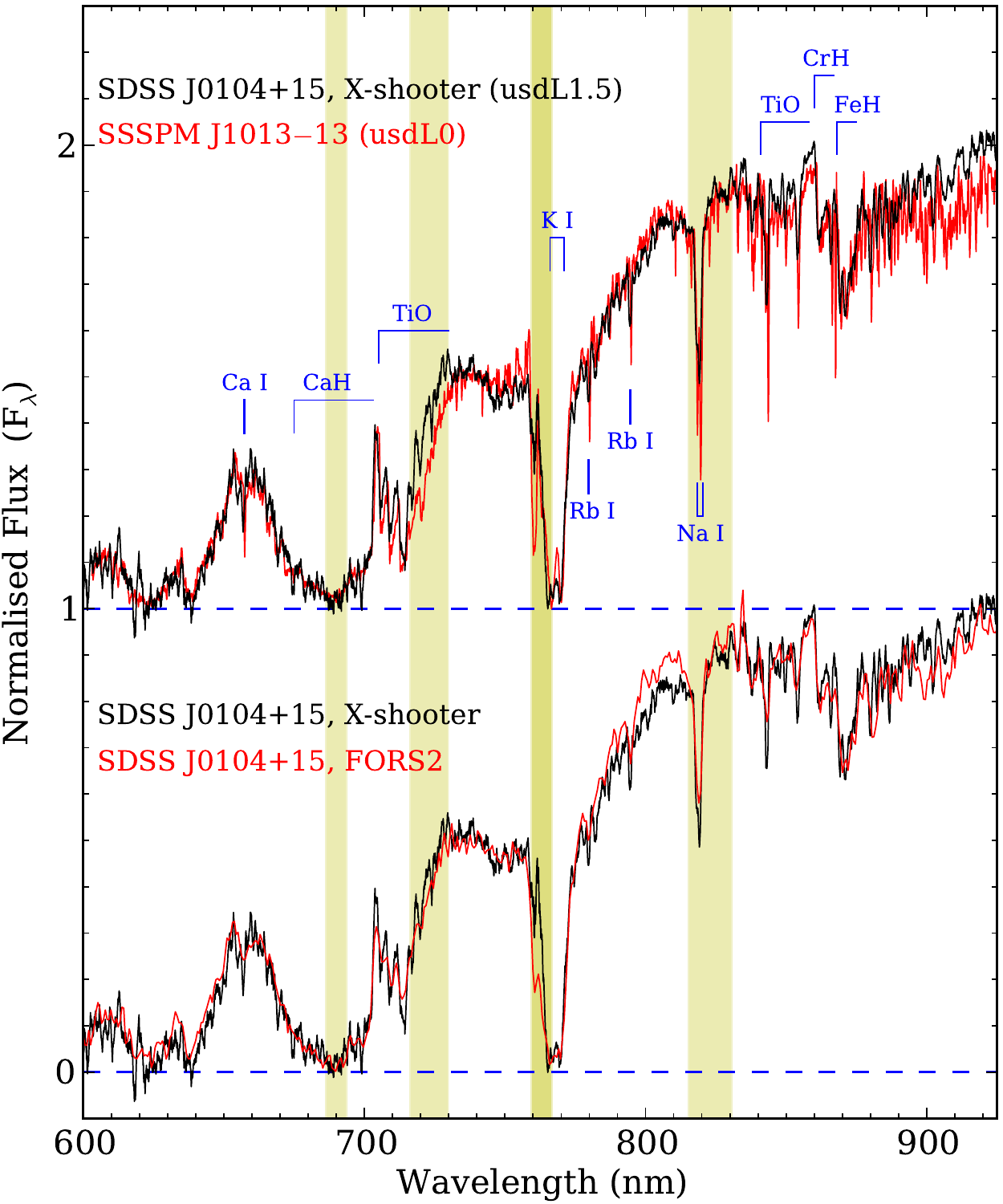}
\caption[]{The optical spectrum of SDSS J0104+15 compared to SSSPM J1013$-$13. The spectrum of SSSPM J1013$-$13 is from \citet[][]{bur07}. The FORS2 spectrum of SDSS J0104+15 is from \citet{lod17}. Spectra are normalized near 860 nm. The spectrum of SDSS J0104+15 was smoothed by a boxcar function of 61 pixels to increase the signal to noise ratio. Telluric absorption regions are highlighted in yellow as in Fig. \ref{spnir}.} 
\label{spvis}
\end{center}
\end{figure}

\begin{figure}
\begin{center}
 \includegraphics[width=\columnwidth]{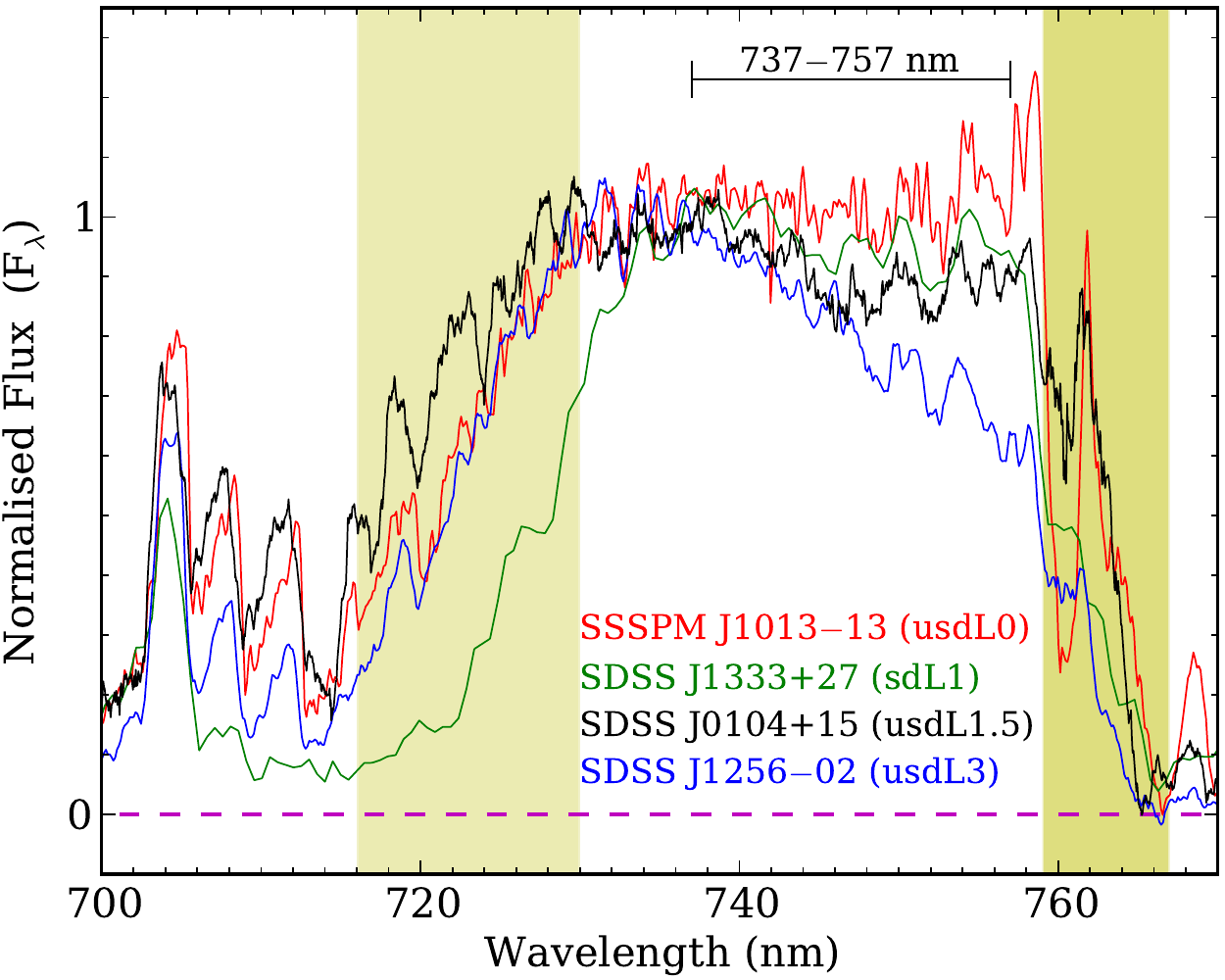}
\caption[]{The 737--757 nm wavelength of SSSPM J1013$-$13, SDSS J0104+15, SDSS J1333+27, and SDSS J1256$-$02 normalized at around 737 nm. The spectrum of SDSS J1256$-$02 is from \citet[][]{bur09}. Telluric absorption regions are highlighted in yellow as in Fig. \ref{spnir}.}
\label{spplat}
\end{center}
\end{figure}

\begin{table}
 \centering
  \caption[]{Properties of SDSS J0104+15.}
\label{prop}
  \begin{tabular}{l l c}
\hline
Parameter & & Value  \\	
\hline
SDSS $\alpha$ (J2000) & .............................. & $01^h04^m48\fs46$ \\
SDSS $\delta$ (J2000) & .............................. & $+15\degr35\arcmin01\farcs8$ \\
SDSS epoch & .............................. & 1999 October13 \\
SDSS $r$ & .............................. & 22.25 $\pm$ 0.17 \\
SDSS $i$ & .............................. & 20.37 $\pm$ 0.05 \\
SDSS $z$ & .............................. & 19.28 $\pm$ 0.06 \\
Pan-STARRS1 $i$    & .............................. & 20.52 $\pm$ 0.02 \\
Pan-STARRS1 $z$    & .............................. & 19.49 $\pm$ 0.02 \\
Pan-STARRS1 $y$    & .............................. & 19.09 $\pm$ 0.03 \\
UKIDSS $Y$ & .............................. & 18.48 $\pm$ 0.05 \\
UKIDSS $J$ & .............................. & 17.93 $\pm$ 0.05 \\
UKIDSS $H$ & .............................. & 18.06 $\pm$ 0.11 \\
UKIDSS $K$ & .............................. & 18.08 $\pm$ 0.17 \\
$WISE ~ W1$ & ..............................  & 16.61 $\pm$ 0.08 \\
$WISE ~ W2$ & .............................. & 16.36 $\pm$ 0.25 \\
Spectral type & .............................. & usdL1.5 $\pm$ 0.5 \\
Distance (pc) & .............................. & 228$^{+61}_{-49}$  \\
$\mu_{\rm RA}$ (mas yr$^{-1}$) & .............................. &  206.2 $\pm$ 4.2 \\
$\mu_{\rm Dec}$ (mas yr$^{-1}$) & .............................. &  $-$179.1 $\pm$ 4.6  \\
$V_{\rm tan}$ (km s$^{-1}$) & .............................. & 276 $\pm$ 75 \\
RV (km s$^{-1}$) & .............................. & $-$26 $\pm$ 16  \\
$U$ (km s$^{-1}$) & .............................. & $-$98 $\pm$ 40 \\
$V$ (km s$^{-1}$) & .............................. & $-$261 $\pm$ 79  \\
$W$(km s$^{-1}$)  & .............................. & $-$100 $\pm$ 46  \\
$T_{\rm eff}$ (K) & .............................. & 2450 $\pm$ 150  \\
$\rm {[Fe/H]}$ & .............................. & $-$2.4 $\pm$ 0.2 \\
$\rm {[M/H]}$ & .............................. & $-$2.1 $\pm$ 0.2 \\
Mass  (M$_{\sun}$) & .............................. & 0.086 $\pm$ 0.0015 \\
Age (Gyr) & ..............................  & 11--13 \\
\hline
\end{tabular}
\end{table}

\begin{figure*}
\begin{center}
 \includegraphics[width=\textwidth]{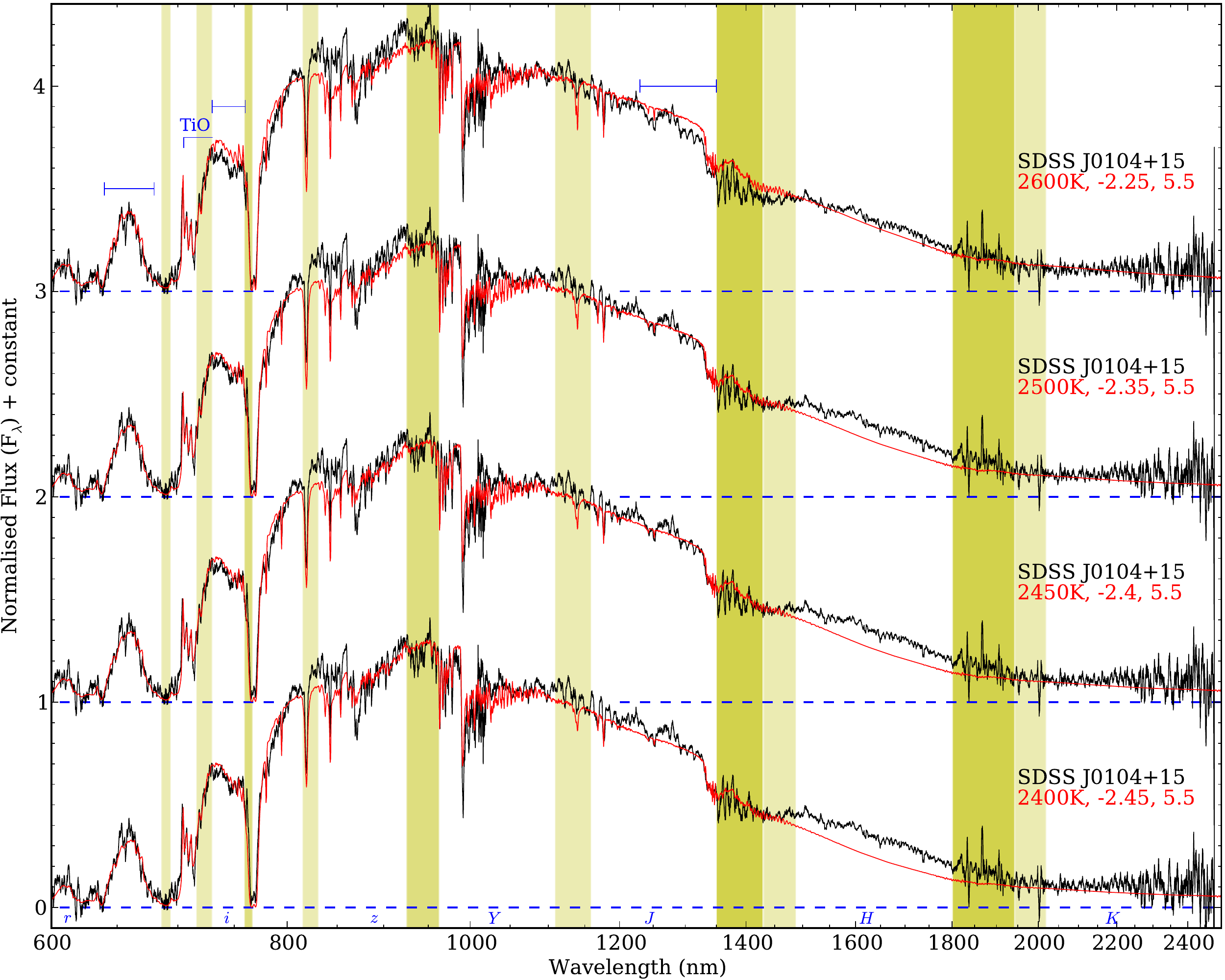}
\caption[]{The optical--NIR spectrum of SDSS J0104+15 compared to BT-Settl model spectra.  The $T_{\rm eff}$, [Fe/H] and log $g$ of model spectra are indicated above their $K$-band spectra. Metallicity and $T_{\rm eff}$ sensitive wavelength ranges (640--680, 705--730, 730--760 and 1230--1350 nm) are marked on the top. The spectrum of SDSS J0104+15 was smoothed by a boxcar function of 61 pixels to increase signal-to-noise ratio. SDSS ($r, i$ and $z$) and UKIDSS ($Z, Y, J, H$ and $K$) filters are marked at their effective wavelengths. Spectra are normalized at 800 nm. The axis tick-marks are spaced logarithmically for clearer display of the optical spectra. Telluric absorption regions are highlighted in yellow same as in Fig. \ref{spnir}.}
\label{spmod}
\end{center}
\end{figure*}

\begin{figure}
\begin{center}
 \includegraphics[width=\columnwidth]{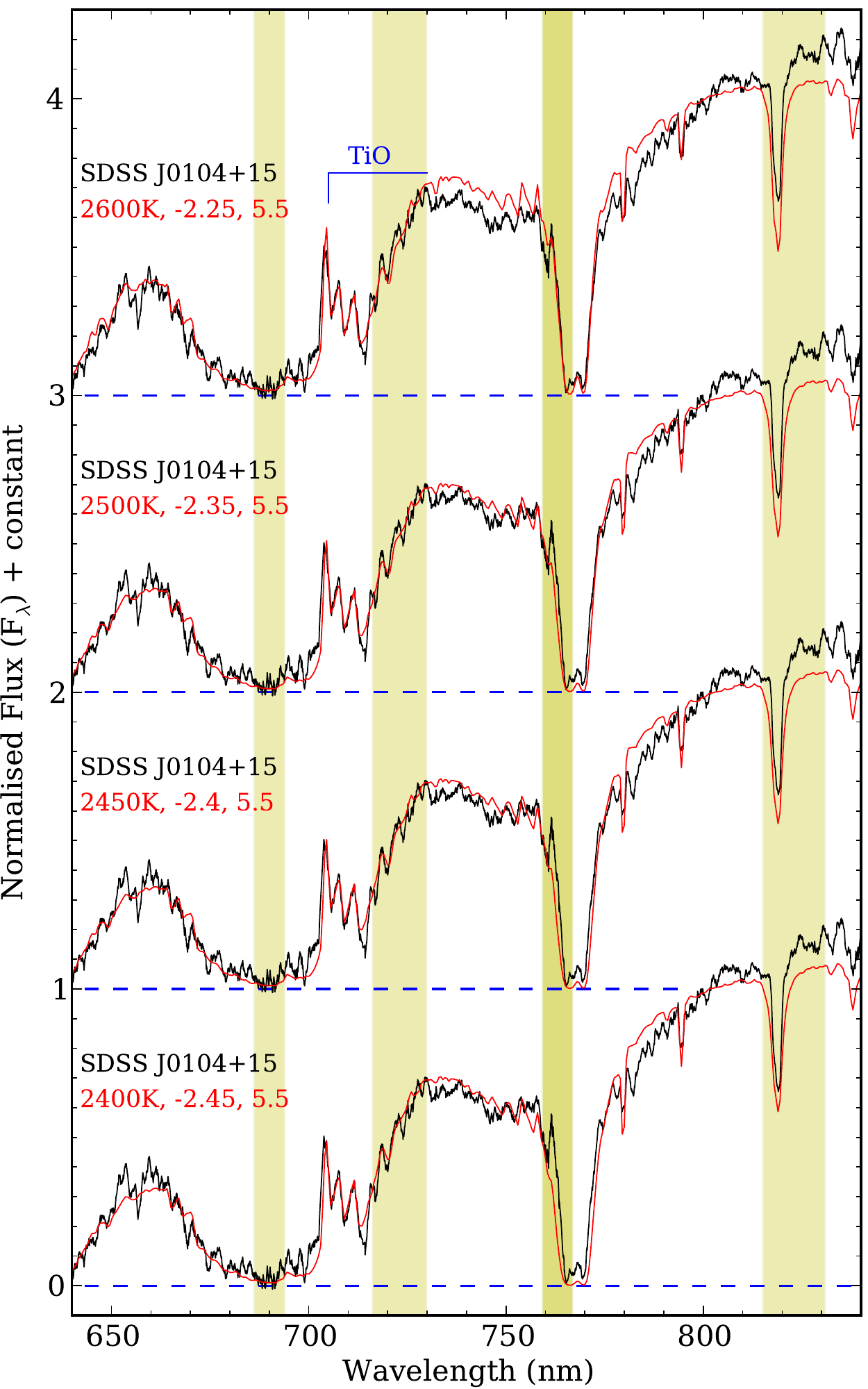}
\caption[]{A zoom in of  Fig. \ref{spmod} at red optical wavelength.  }
\label{sptio}
\end{center}
\end{figure}

\section{Characteristics}
\label{spro}
\subsection{Spectral classification}
\label{ssc}

Fig. \ref{spnir} shows the new optical--NIR spectrum of SDSS J0104+15 compared to that of a usdL0 subdwarf \citep[SSSPM J1013$-$13;][]{bur04,sch04,zha17}. SDSS J0104+15 has stronger overall suppression in the NIR as well as a flatter $K$-band morphology, both of which can be accounted for (according to the model atmospheres) by stronger enhanced collision-induced H$_2$ absorption \citep[CIA H$_2$;][]{bat52,sau12}. This is consistent with SDSS J0104+15 being more metal-poor than SSSPM J1013$-$13. Fig. \ref{spvis} shows only the optical spectrum of SDSS J0104+15 compared to that of SSSPM J1013-13. These objects have similar optical spectral profiles, however SDSS J0104+15 has weaker TiO absorption bands at around 710 and 850 nm, offering further evidence that SDSS J0104+15 is lower metallicity than SSSPM J1013$-$13. Therefore, SDSS J0104+15 is likely an early type usdL subdwarf.

The slope of the spectra at 737--757 nm wavelength is used to assign spectral types of early L subdwarfs \citep{kirk14,zha17}. In the 737--757 nm range, the slope of the spectrum is positive (i.e., the spectrum is red) for L0, flat for L0.5, and negative for L1 and later types \citep[see fig. 10 in][]{zha17}. In the 737--757 nm wavelength range, the slope of the spectra of early L-type objects is bluer at both lower [Fe/H] and $T_{\rm eff}$. Therefore, a esdL0.5 type spectrum has a higher $T_{\rm eff}$ than an sdL0.5 type spectrum. Meanwhile, a usdL subclass spectrum has a later subtype than an sdL subclass spectrum with the same $T_{\rm eff}$. For instance, a usdL2 type object would have similar $T_{\rm eff}$ as an sdL0 type object \citep[see fig. 20 in][]{zha17}. Fig. \ref{spplat} compares the 737--757 nm wavelength range in the spectrum of SDSS J0104+15 to those of SSSPM J1013-13 (usdL0), SDSS J133348.24+273508.8 \citep[SDSS J1333+27, sdL1;][]{zha17}s SDSS J1256$-$02 (usdL3), with the spectra normalized at around 737 nm. The slope of the spectrum of SDSS J0104+15 in the 737--757 nm wavelength range is approximately intermediate between the usdL0 and usdL3 comparison objects, and slightly bluer than the sdL1. The slope is clearly much closer to that of the sdL1 than to the usdL3, so we chose a spectral classification of usdL1.5 $\pm$ 0.5 for SDSS J0104+15. 

In retrospect we note that SDSS J0104+15 (usdL1.5) and the earlier usdL0 SSSPM J1013$-$13 have similar flux ratios between 740 and 810 nm, despite this ratio increasing with increasing $T_{\rm eff}$. However, this ratio is also sensitive to metallicity, increasing with decreasing [Fe/H]. So the 740 to 810 nm similarity could be explained if SDSS J0104+15 has lower metallicity and cooler $T_{\rm eff}$ compared to SSSPM J1013$-$13. This will be discussed further in Section \ref{sap}.

\subsection{Halo kinematics}
\label{shk}
We derived spectroscopic distance estimates for SDSS J0104+15 using the relationship between spectral type and $J$- and $H$- band absolute magnitude shown in fig. 16 of \citet{zha17}. We obtained distance constraints of $215^{+44}_{-36}$ pc and  $241^{+49}_{-41}$ pc in the $J$ and $H$ bands, respectively. We adopt the average distance estimate and uncertainty of these $J$ and $H$ band estimates, giving $228^{+61}_{-49}$ pc. We estimated the $Gaia ~ G$- band magnitude of SDSS J0104+15 to be 20.93 $\pm$ 0.21 using the relationship between $G - r$ and $r - i$ colours \citep{jord14}. This is close to the $Gaia$ limit \citep[$G \simeq 20.7$;][]{gaia16}s SDSS J0104+15 is thus a borderline $Gaia$ object. It may be detected by $Gaia$ in its final data release, but with a somewhat lower parallax accuracy compared to brighter ($G$ < 20) objects.

The proper motion of SDSS J0104+15 was measured from SDSS $i$ and PS1 $i_{\rm P1}$-band images which have a baseline of 13.2 yr. We used the {\scriptsize IRAF} task {\scriptsize GEOMAP} to derive spatial transformations from the SDSS $i$ into the PS1 $i_{\rm P1}$-band image.  Thirteen reference stars around SDSS J0104+15 were used for the transformation. These transforms allowed for linear shifts and rotation. We then transformed the SDSS pixel coordinates of SDSS J0104+15 into the PS1 image using {\scriptsize GEOXYTRAN}, and calculated the change in position (relative to the reference stars) between the two epochs. This analysis yield $\mu_{\rm RA} = 206.2\pm4.2$ mas yr$^{-1}$ and $\mu_{\rm Dec} = -179.1\pm4.6$ mas yr$^{-1}$. The errors on proper motion are computed from the root mean square of the position shifts of reference stars between SDSS and PS1 fields. 

To facilitate RV determination for SDSS J0104+15 we obtained an X-shooter spectrum of an L1 dwarf \citep[DENIS-P J1441$-$0945;][]{mart99} with known RV \citep[$-$27.9 $\pm$ 1.2 km s$^{-1}$;][]{bail04}. We then cross correlated strong absorption lines (Rb I, Na Is K I) in the optical and NIR between SDSS J0104+15 and DENIS-P J1441$-$0945. The RV of SDSS J0104+15 was found to be $-$85 $\pm$ 6 km s$^{-1}$. The RV error is from the standard deviation of RV measurements from different absorption lines. 

The Galactic $UVW$ space motions of SDSS J0104+15 were determined using our spectroscopic distance, RV and proper motion following \citet{clar10}. It has typical halo velocities: $U = -98\pm40$ km s$^{-1}$,  $V = -261\pm79$ km s$^{-1}$ and $W = -100\pm46$ km s$^{-1}$ [see fig. 17 of \citet{zha17} for comparison; here $U$ is positive in the direction of the Galactic anti centre, $V$ is positive in the direction of Galactic rotations $W$ is positive in the direction of the North Galactic Pole \citep{joh87}]. Table \ref{prop} summarises the properties of SDSS J0104+15. 

\subsection{Atmospheric properties}
\label{sap}
We used the BT-Settl models \citep{alla14} to constrain the atmospheric parameters of SDSS J0104+15. The BT-Settl atmospheric models can reproduce the overall observed spectra of M and L subdwarfs, and can closely reproduce a variety of optical and NIR spectral features. BT-Settl models are able to reproduce observed spectra rather better for M and L subdwarfs with [Fe/H] < $-$1.0 than for [Fe/H] > $-$1.0 \citep{zha17}. 

 The model grids we used cover 2000 K $\leq T_{\rm eff} \leq$ 2600 K, $-2.5 \leq$ [Fe/H] $\leq -0.5$ and 5.0 $\leq$ log $g$ $\leq$ 5.75, with intervals of 100 K for $T_{\rm eff}$, 0.5 dex for [Fe/H], and 0.25 dex for log $g$, and account for $\alpha$-enhancement ([$\alpha$/Fe] = +0.4 is adopted for [Fe/H] $\leq --$1.0, and [$\alpha$/Fe] = +0.2 is adopted for [Fe/H] = $-$0.5).  The relation between [M/H] and [Fe/H] is [M/H] $\approx$ [Fe/H] + 0.3 for scaled solar compositions with [$\alpha$/Fe] = +0.4, and [M/H] $\approx$ [Fe/H]+0.15 for  [$\alpha$/Fe] = +0.2. We used linear interpolation between some models if this yielded a significantly improved fit. 

Surface gravity has the least effect on the spectral profile of L subdwarfs compared to temperature and metallicity. \citet{zha17} has shown that esdM7--esdL4 subdwarfs have a similar log $g$ of $\sim$5.5 dex, with their spectra being mainly affected by $T_{\rm eff}$ and metallicity. Therefore, we used model spectra with log $g$ = 5.5 dex for our comparisons with SDSS J0104+15 to find the closest model-fit $T_{\rm eff}$ and [Fe/H]. While the BT-Settl models can reasonably reproduce the overall spectral profile of early L dwarfs, some detailed features are not reproduced that well \citep{zha17}. Furthermore, some wavelength ranges are more sensitive to $T_{\rm eff}$ and/or [Fe/H] than others. We therefore performed a by-eye comparison between model spectra and SDSS J0104+15, focusing on a set of sensitive well modelled wavelength regions.

The 640--680 nm wavelength region and TiO absorption band at 705--730 nm are particularly sensitive to [Fe/H] for early type L subdwarfs with [Fe/H] < $-$2.0 \citep[see fig. 6 of][]{zha17}. The 705--730 nm TiO absorption band is weakening rapidly from [Fe/H] = $-$2.0 to $-$2.5, and responses to small changes of [Fe/H] (e.g. 0.05 dex). Also the 730--760 nm wavelength region is very sensitive to [Fe/H] and $T_{\rm eff}$ \citep[figs 10 and 13 of][]{zha17}. We followed a two-step approach for our by-eye model-fitting. First we considered the 705--730 nm TiO absorption band and the 640--680 nm wavelength region, and identified a set of good-fitting models with $T_{\rm eff} \lid$ 2600 K (with step sizes of 50 K on $T_{\rm eff}$ and 0.05 dex on [Fe/H]). We then compared this good-fitting model set to the 730--760 nm wavelength and the NIR regions, and further refined our best-fitting model selection to obtain $T_{\rm eff}$ and [Fe/H] constraints. 

Fig. \ref{spmod} shows the optical--NIR spectrum of SDSS J0104+15 compared to our four good-fitting models in the 710 nm TiO absorption band and the 640-680 nm wavelength region. Four model spectra all fit well with the overall spectral profile of SDSS J0104+15. Fig. \ref{sptio} shows a zoom-in of Fig. \ref{spmod} at 640--840 nm.  The 2600 K model spectrum has more flux at 730--760 nm than SDSS J0104+15, and does not fit well with the steep shoulder at 1227 nm, which is also sensitive to $T_{\rm eff}$.  Therefore, SDSS J0104+15 should have a $T_{\rm eff}$ below 2600 K. The other three model spectra have a bit less flux at $H$ band, however they fit well with SDSS J0104+15 at these metallicity and $T_{\rm eff}$ sensitive regions from 600 to 1350 nm, and thus constitute our refined best-fitting model selection. In this way, we estimate that SDSS J0104+15 has $T_{\rm eff}$ = 2450 $\pm$ 150 K and [Fe/H] = $-$2.4 $\pm$ 0.2, accounting for possible systematic uncertainties. The total metallicity of SDSS J0104+15 is [M/H] = $-$2.1 $\pm$ 0.2. SDSS J0104+15 would have an age of 11--13 Gyr according to ages of stars with similar metallicity in globular clusters and the Galaxy's halo \citep{jofr11,dott10}.

We can now compare the observed colours of SDSS J0104+15 directly to model predictions using Fig. \ref{ijk}, which shows the $i-J$ and $J-K$ colours calculated for model atmospheres with $T_{\rm eff}$ of 2000--5000 K, $-2.5 \leq$ [Fe/H] $\leq -0.5$, and log $g$ of 5.5. The best-fitting models predict a bluer $J-K$ colours than SDSS J0104+15. We suggest that the detailed continuum shape of the BT-Settl model spectra could still be improved in this VMP domain ([Fe/H] < $-$2.0).

\begin{figure}
\begin{center}
  \includegraphics[width=\columnwidth]{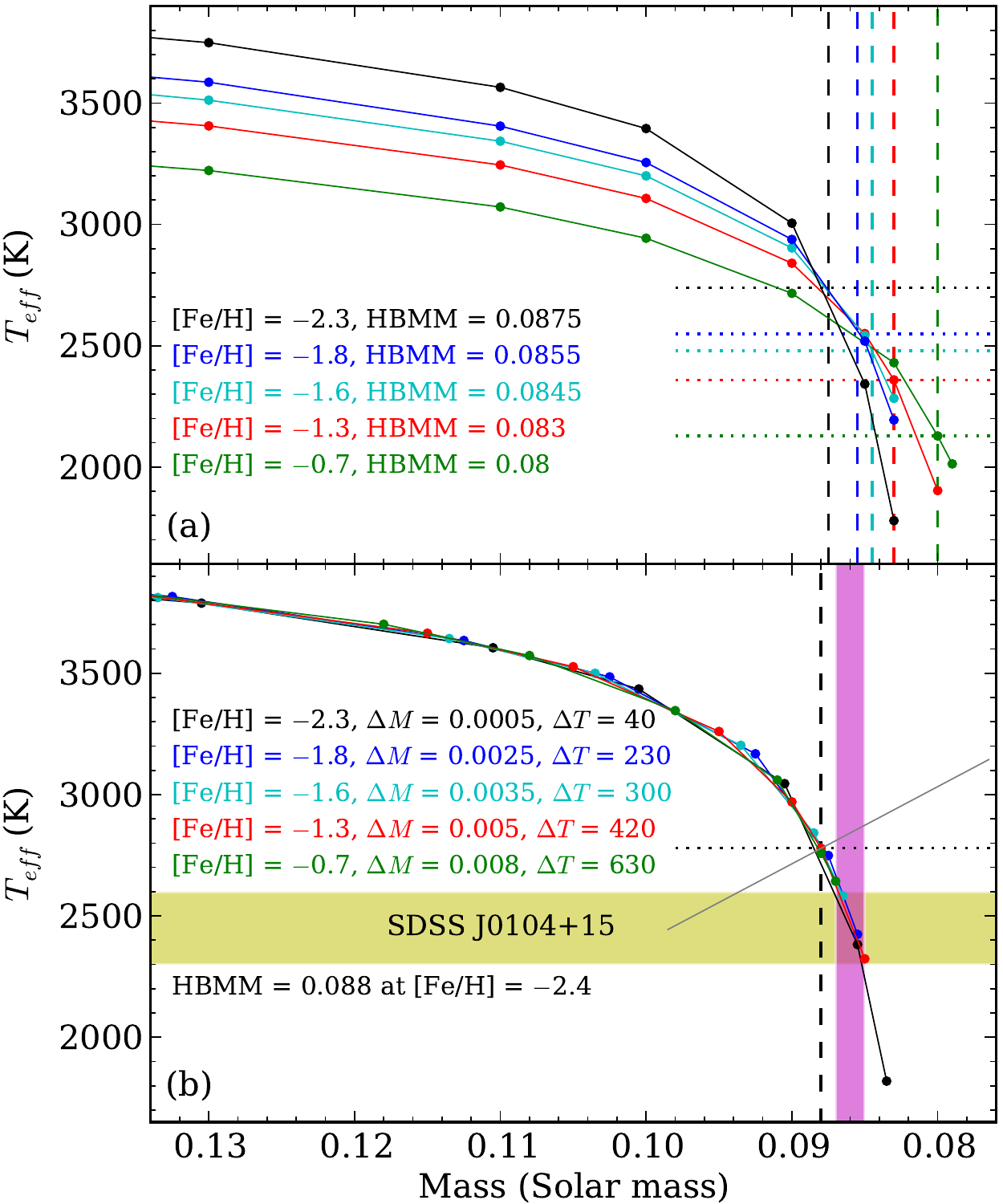}
\caption[]{(a) The mass--$T_{\rm eff}$ relationships at 10 Gyr derived from evolutionary models \citep{bara97}. Black, blue, red and green vertical lines indicate the HBMMs at [M/H] of $-$2.0, $-$1.5,  $-$1.0, and $-$0.5. (b) The relationships for [M/H] of $-$1.5, $-$1.0, and $-$0.5 were shifted along mass and $T_{\rm eff}$ axes to match with the profile of [M/H] = $-$2.0. Shifted values are labeled on the plot. }
\label{mteff}
\end{center}
\end{figure}

\begin{figure*}
\begin{center}
   \includegraphics[width=\textwidth]{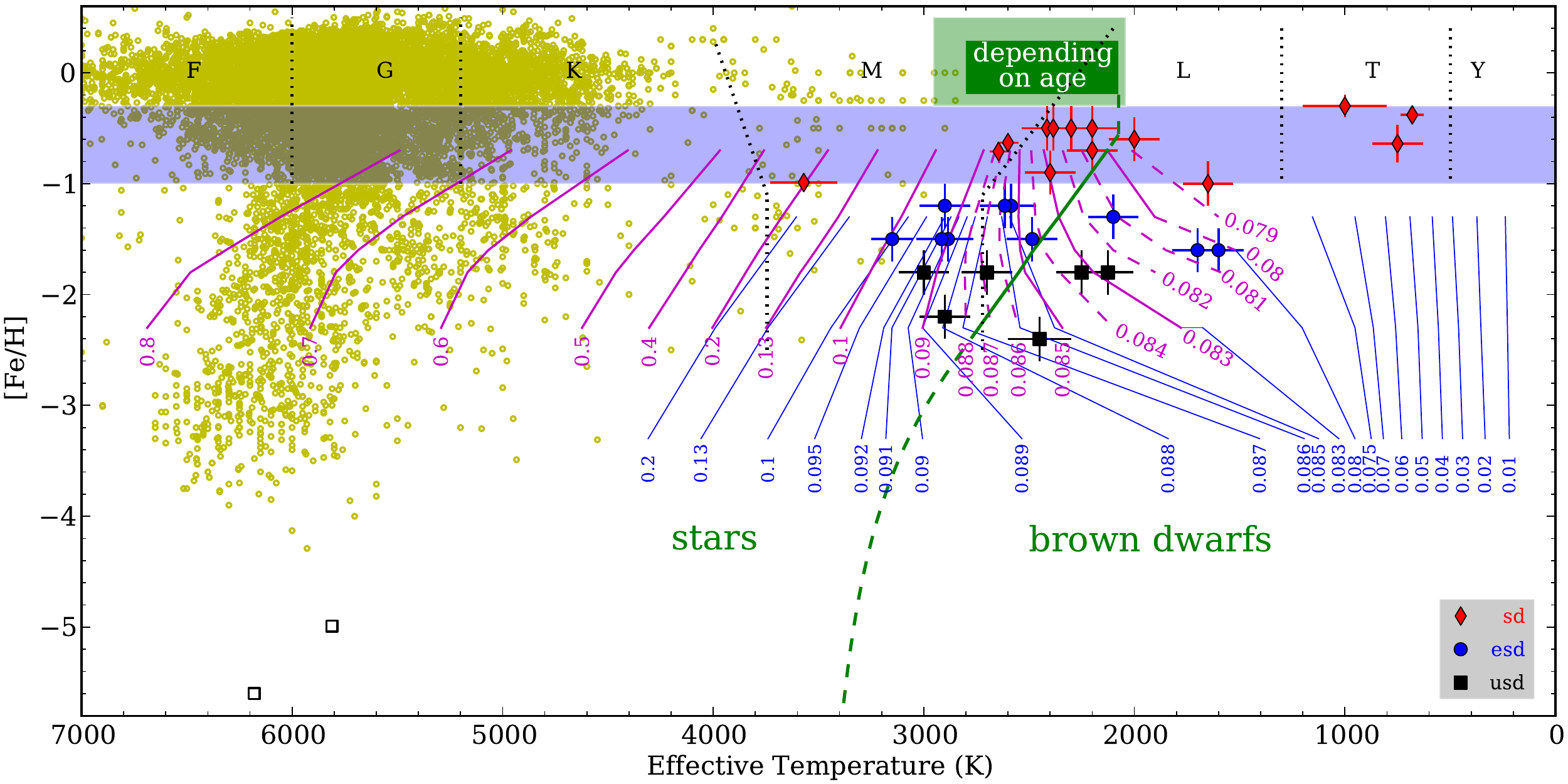}
\caption[]{[Fe/H] and $T_{\rm eff}$ of cool and ultra-cool subdwarfs. The shaded blue area indicates the approximate [Fe/H] range for the thick disc population \citep[e.g.][]{spag10}, with the thin disc population above and the halo population below. Black dotted lines indicate the boundaries between F, G, K, M, L, T, and Y types.  Magenta lines indicate the 10 Gyr iso-mass contours \citep{cha97,bara97} with mass values (in M$_{\sun}$) marked below or next to each iso-mass line. The green solid line indicates the $T_{\rm eff}$ of the HBMM at $-2.3 \lid$ [Fe/H] $\lid -0.7$. Shaded green area is where both VLMS and massive brown dwarfs could appear depending on age. Blue iso-mass contour lines are based on calculations of \citet{burr98}. SDSS J0104+15 is the filled black square at [Fe/H] = $-$2.4. Yellow open circles are dwarf stars (log $g$ > 3.5) from the PASTEL catalogue \citep{soub16}. The red diamond near the 0.2 M$_{\sun}$ iso-mass contour is Kapteyn's star (sdM1) measured by \citet{wool05}. Two black open squares are from \citet{freb05} and \citet{caff11}. [Fe/H] measurements of two late type sdM, and three sdT subdwarfs come from their primary stars \citep{bow09,agan16,murr11,pin12,mace13}. The esdM object on the 0.1 iso-mass line is a companion to a K subdwarf \citep{pavl15}. The remaining late type subdwarfs are from \citet{zha17}.  Note the $T_{\rm eff}$ of some objects are offset by $\pm$15 K for clarify when they share the same $T_{\rm eff}$ and  [Fe/H].}
\label{mostmp}
\end{center}
\end{figure*}

\subsection{The hydrogen-burning minimum mass}

The central temperature ($T_{\rm c}$) of VLMS with 0.1--0.3 M$_{\sun}$ is independent of metallicity. Fig. 6 of \citet{cha97} shows that the mass--$T_{\rm c}$ relationships at 0.1--0.3 M$_{\sun}$ are the same for [M/H] = 0 and $-$1.5. The lower the metallicity, the lower the opacity and the more transparent the atmosphere, and the same optical depth lies at deeper layers with higher temperature in more metal-poor stars. Therefore, more metal-poor stars have higher $T_{\rm eff}$ than metal-rich stars with same mass. However, a 10 Gyr metal-poor brown dwarf could have cooler $T_{\rm eff}$ than a metal-rich star with the same mass. This is because the HBMM is lower at higher metallicity than at lower metallicity, and the higher metallicity means higher opacity, which in turn produces higher $T_{\rm c}$ by steepening the temperature gradient. For the same reason, a massive metal-poor brown dwarf could have the same mass as a least massive metal-rich star \citep{burr01}.  

Evolutionary models show that nuclear ignition still takes place in the central part of stars with mass slightly below $\sim$0.083 M$_{\sun}$ at [M/H] = $-$1.0, but cannot balance steadily the ongoing gravitational contraction, which defines the massive brown dwarfs \citep{cha97}. The same occurs  in stars with mass slightly below $\sim$ 0.072 M$_{\sun}$ at [M/H] = 0. Therefore, the HBMMs are $\sim$ 0.072 M$_{\sun}$ at [M/H] = 0 and $\sim$ 0.083 M$_{\sun}$ at [M/H] = $-$1.0. The exact HBMM at [M/H] < $-$1.0 is not explicitly stated in \citet{cha97} and \citet{bara97}. The HBMM at primordial metallicity ($Z = 0$) is $\sim$ 0.092 M$_{\sun}$ according to \citet{burr01}. In this section we use the mass--$T_{\rm eff}$ relations given by evolutionary models to try to deduce the HBMM at various metallicities. 

Fig. \ref{mteff} (a) shows the 10 Gyr mass--$T_{\rm eff}$ relationships derived from evolutionary models of low-mass stars with [M/H] of $-$0.5, $-$1.0, $-$1.3, $-$1.5, and $-$2.0 \citep{cha97,bara97}. Note that the [M/H] scale is not calibrated for $\alpha$-enhancement.  We converted the [M/H] to [Fe/H] scale with [M/H] = [Fe/H]+0.3 ([$\alpha$/Fe] = +0.4) for [M/H] $\lid$ $-$1.0 and [M/H] = [Fe/H]+0.2 ([$\alpha$/Fe] = +0.3) for [M/H] = $-$0.5. These evolutionary models  employed the base atmospheric models of \citet{alla95}. The steepening of the mass--$T_{\rm eff}$ relationship near the lower mass end reflects the onset of ongoing electron degeneracy in the stellar interior, which is the characteristic of the transition between the stellar and sub-stellar domains. $T_{\rm eff}$ is a decreasing function of metallicity above the HBMM, but an increasing function of metallicity below the HBMM.  A mass--$T_{\rm eff}$ relationship at a certain  [Fe/H] intersects with other relationships at different  [Fe/H]. The intersection points with the relationships at higher [Fe/H] provide upper limits on the HBMM at the certain [Fe/H]. For example, the mass--$T_{\rm eff}$ relationships at [Fe/H] = $-$1.3 and [Fe/H] = $-$0.7 intersect around 0.084 M$_{\sun}$. Therefore, the HBMM at [Fe/H] = $-$1.3 is expected to be below 0.084 M$_{\sun}$.

Fig. \ref{mteff} (b) shows mass--$T_{\rm eff}$ relationships that have been shifted along mass and $T_{\rm eff}$ axes to best match with each other, at a projected position at [Fe/H] = $-$2.4. We shifted these relationships with steps of 0.0005 M$_{\sun}$ and 10 K. These shifted final values of mass (in M$_{\sun}$) and $T_{\rm eff}$ (in K) are indicated on the plot. These relationships of different [Fe/H] have very similar profiles at 0.08--0.3 M$_{\sun}$. This is likely because that the mass--$T_{\rm c}$ and mass--radius relationships at 0.1--0.3 M$_{\sun}$ are very similar at different metallicity, and the steepening of the mass--$T_{\rm eff}$ relationship near the lower-mass end are caused by the same physical reason, which is electron degeneracy in the stars at stellar-substellar transition. Therefore, the cross points of HBMMs on these relationships at different [Fe/H] are overlapped in Fig. \ref{mteff} (b). The perpendicular line at the HBMM on these relationships is marked in Fig. \ref{mteff} (b). The mass shift of a relationship at a certain [Fe/H] to match the relationship profile at [Fe/H] = $-$1.3, is also the HBMM shift relative to the HBMM at [Fe/H] = $-$1.3, which is 0.083 M$_{\sun}$. Therefore, the HBMMs are 0.0875, 0.0855, 0.0845, 0.083, and 0.08 M$_{\sun}$ at [Fe/H] = $-$2.3, $-$1.8, $-$1.6, $-$1.3, and $-$0.7, respectively, according to Fig. \ref{mteff} (b). The corresponding $T_{\rm eff}$ at 10 Gyr are 2739, 2549, 2479, 2359, and 2128 K, respectively. The HBMMs and $T_{\rm eff}$ at these five  [Fe/H] values are indicated as vertical dashed lines and horizontal dotted lines in Fig. \ref{mteff} (a), respectively. The projected HBMM at [Fe/H] = $-$2.4 is around 0.088 M$_{\sun}$. SDSS J0104+15 has a $T_{\rm eff }$ = 2450 $\pm$ 150 K, indicated with the shaded-yellow belt in Fig. \ref{mteff} (b). The corresponding mass of  SDSS J0104+15 derived from the mass--$T_{\rm eff}$ relationship at  [Fe/H] = $-$2.4  is between  0.085 and 0.087 M$_{\sun}$, which is indicated with a shaded-magenta belt. The mass uncertainty caused by $T_{\rm eff}$ error (150 K) is 0.001 M$_{\sun}$. The mass uncertainty caused by [Fe/H] error (0.2 dex) is around 0.008--0.001 M$_{\sun}$, as  the mass--$T_{\rm eff}$ relationship at [M/H] = $-$1.5 was shifted by 0.002 M$_{\sun}$ to match with the relationship at [M/H] = $-$2.0 (Fig. \ref{mteff} b). Age uncertainty may affects our mass estimation by up to 0.0005 M$_{\sun}$. Because the $T_{\rm eff}$ of a massive brown dwarf drop by $\sim$ 50--100 K from 10 Gyr to 11--13 Gyr \citep[e.g.][]{bara03}.  The square root of the sum of squares of all uncertainties is 0.0015 M$_{\sun}$.  Therefore, SDSS J0104+15 has a mass of  0.086 $\pm$ 0.0015 M$_{\sun}$. 

Fig. \ref{mostmp} explores how the most metal-poor subdwarf population distribution maps on to the [Fe/H]--$T_{\rm eff}$ plane for F, G, K, M, L and T types. 10 Gyr iso-mass contour lines are plotted to better visualize the HBMM at different [Fe/H].
Solid magenta contour lines are from \citet{cha97} and \citet{bara97}. We also show some interpolated contours (dashed magenta lines) based on mass--$T_{\rm eff}$ relationships at different metallicity (which have very similar profiles; see Fig. \ref{mteff} b). Blue contour lines are from \citet{burr98}, and will further aid discussion in Section 3.5. 
Guided by these model contour lines we have generated a HBMM limit in the [Fe/H]--$T_{\rm eff}$ plane over the range $-2.3 \lid$ [Fe/H] $\lid -0.7$, which is shown as a solid green line that is well approximated by the straight line function: 
\begin{equation}
T_{\rm eff} = 1861 - 382 \times {\rm [Fe/H]}
\end{equation}
A green box area indicates the overlapped $T_{\rm eff}$ region for young brown dwarfs and older VLMS in the solar neighbourhood. VLMS just above the HBMM have $T_{\rm eff} \ga$ 2075 K \citep{diet14}. Meanwhile, PPl 15 AB \citep{basr96}, a young binary brown dwarf confirmed by the lithium test \citep{maga93} in the Pleiades open cluster, has a $T_{\rm eff}$ of 2800 $\pm$ 150 K \citep{rebo96}.
The corresponding $T_{\rm eff}$ of the HBMM ($\sim$0.092 M$_{\sun}$) at primordial metallicity is $\sim$3600 K \citep{burr01}. We have thus extended our HBMM line to lower metallicity ([Fe/H] < $-$2.3) following a tangent function. This extended (green dashed) line approaches 3600 K at [Fe/H] = $-\infty$, and is described by:
\begin{equation}
 {\rm [Fe/H]} = -2.3 -1.43 \times \tan\frac{T_{\rm eff} - 1017}{548} 
\end{equation}
The corresponding 10 Gyr $T_{\rm eff }$ at [Fe/H] =$-$2.4 is around 2777 K according to equation (2).
We also conservatively extend the HBMM line to higher metallicity by joining it on to the right side of the green box, which provides a reference of $T_{\rm eff}$ for the HBMM at [Fe/H] >  $-$0.7.
It can be seen that the 10 Gyr iso-mass lines for 0.085 and 0.083 M$_{\sun}$ turn to cooler $T_{\rm eff}$ below the HBMM limit at   [Fe/H] = $-$1.7 and [Fe/H] = $-$1.3, respectively. This is consistent with the steep $T_{\rm eff}$ decent in the mass--$T_{\rm eff}$ relationship below the HBMM, that is seen at different metallicities in Fig. \ref{mteff}.

SDSS J0104+15 is clearly on the substellar side of the HBMM limit, and according to our analysis joins five other halo L subdwarfs that are brown dwarfs; 2MASS J1626+39, SDSS J1256$-$02, ULAS J151913.03$-$000030.0 \citep[ULAS J1519$-$00;][]{zha17}, 2MASS J06164006$-$6407194 \citep[2MASS J0616$-$64;][]{cus09}, and 2MASS J05325346+8246465 \citep[2MASS J0532+82;][]{bur03}. SDSS J0104+15 appears to be the most metal-poor brown dwarf identified to-date, and is also the most massive brown dwarf yet known. 

To aid early identification of metal-poor brown dwarfs we have transferred our stellar--substellar boundary line on to the $i-J$ versus $J-K$ colour-colour diagram, based on the observed colours of SDSS J0104+15 and the other objects with constrained $T_{\rm eff }$ and [Fe/H]  \citep[from][]{zha17} in Fig. \ref{mostmp}. This approximate stellar--substellar boundary is indicated in Fig. \ref{ijk} as a black dashed line.

\subsection{The halo brown dwarf transition zone}

Returning to Fig. \ref{mostmp} the 10 Gyr iso-mass contours of \citet[][blue lines]{burr98} span a very interesting region of the metallicity--$T_{\rm eff}$ plane. These models were calculated across 0.01--0.2 M$_{\sun}$ at $Z$ = 0.1, 0.01, and 0.001 Z$_{\sun}$ (i.e. [Fe/H] = $-$1.3, $-$2.3, and $-$3.3, respectively; [$\alpha$/Fe] = +0.4 is adopted), with base atmospheric models from \citet{alla95}. Each of these contour lines has three data points at  [Fe/H] = $-$1.3, $-$2.3, and $-$3.3, and we note that the  0.08 and 0.083 M$_{\sun}$ iso-mass lines join almost seamlessly on those of \citet{cha97} (with differences of only $\sim$10 K in $T_{\rm eff}$ at  [Fe/H] = $-$1.3). The mass--$T_{\rm eff}$ relationship (in the range 0.01--0.2 M$_{\sun}$) shown by the Burrows models \citep[e.g. fig 5;][]{burr01} leads to a `transition zone' below the HBMM and above $T_{\rm eff} \approx$ 1200 K, where object $T_{\rm eff}$ is very sensitive to mass and metallicity. The internal energy of halo brown dwarfs in this transition zone is partially provided by unsteady nuclear fusion \citep[e.g. fig. 8;][]{cha97}. This transition zone is also manifest as a substellar subdwarf gap between the $T_{\rm eff}$ evolutionary tracks of low-mass stars and brown dwarfs \citep[e.g. fig. 8;][]{burr01}, which should lead to a sparsity of objects in this region \citep[e.g. fig. 10;][]{burg04} due to the narrow mass range across a broad $T_{\rm eff}$.

The transition zone region is clear in our Fig. \ref{mostmp}, lying between the green HBMM limit and $T_{\rm eff} \approx$ 1200 K. The width of the $T_{\rm eff}$ range of the transition zone increase from $\sim$ 1000 K at [Fe/H] = $-$1.0 to $\sim$ 1800 K at [Fe/H] = $-$3.3. Most of the esdL and usdL subdwarfs are in the transition zone except for some early type L subdwarfs that are VLMS just above the HBMM. SDSS J0104+15, 2MASS J1626+39, SDSS J1256$-$02, ULAS J1519$-$00, 2MASS J0616$-$64, and 2MASS J0532+82 are all in the  transition zone.

Halo brown dwarfs with mass of $\sim$ 0.075--0.01 M$_{\sun}$ should have evolved to T and Y types after over $\sim$ 10 Gyr of cooling. However, we have not found such objects to-date (with expected $T_{\rm eff} \la 1200$ K and [Fe/H] $\la -$1.0). T and Y dwarfs have significantly higher number density in the solar neighbourhood \citep[e.g. fig. 11;][]{kirk12}. If the dependence of substellar formation on metallicity is negligible \citep[as suggested by numerical simulations;][]{bate14}, the ratio between T/Y and L subdwarfs in the halo should be much higher than that of T/Y and L dwarfs, since old halo L subdwarfs cover a much narrower mass range. This points towards a large population of undiscovered T and Y subdwarfs in the local volume.

\section{Summary and conclusions}
\label{ssac}

We have presented an X-shooter optical--NIR spectrum of SDSS J0104+15, and re-classified this object as a usdL1.5 subdwarf. We measured its astrometry and kinematics and determined $T_{\rm eff}$ and [Fe/H] by fitting the spectrum to the BT-Settl models. With [Fe/H]=$-$2.4 $\pm$ 0.2 SDSS J0104+15 is the most metal-poor L subdwarf known to-date. We also constructed a metallicity--$T_{\rm eff}$ diagram, within which we identified the location of the HBMM limit and a halo brown dwarf transition zone beneath this limit down to $\sim$1200 K. This transition zone is caused by a steep $T_{\rm eff}$ decline in the mass--$T_{\rm eff}$ relationships across the stellar--substellar boundary, due to unsteady nuclear fusion. It covers a narrow mass range but spans a wide $T_{\rm eff}$ range, leading to a substellar subdwarf gap over the mid L to early T type range. Our $T_{\rm eff}$ and [Fe/H] estimates for SDSS J0104+15 place it below the HBMM boundary making it the most metal-poor (and highest mass) brown dwarf yet known. Joining SDSS J0104+15 in the transition zone we identify 2MASS J0532+82, 2MASS J0616$-$64, SDSS J1256$-$02, 2MASS J1626+39, and ULAS J1519$-$00. The existence of substellar objects that are as metal poor as SDSS J0104+15 supports formation theories for stars in this mass and metallicity domain \citep{clar11,grei11,basu12,bate14}.

Large scale NIR surveys, such as the `Visible and Infrared Survey Telescope for Astronomy' \citep[VISTA;][]{suth15} Hemisphere Survey \citep[VHS;][]{mcma13} have great potential to identify additional objects that are more metal poor and cooler than SDSS J0104+15. Improvements in ultra-cool model atmospheres will guide future searches for VMP VLMS and brown dwarfs. Accurate theoretical predictions of $H$-band flux are particularly important, because it is more difficult to detect these objects in the $K$ band that is largely suppressed due to enhanced CIA H$_2$. Further more, the future ESA $Euclid$ \citep{laur11} spectroscopic survey covers a wavelength range of 1100-2000 nm (approximately covering the $J$ and $H$ bands), and information from $H$ band spectra will be very important for the characterization of these objects with $Euclid$.

\section*{Acknowledgements}
This work is based in part on data obtained as part of the UKIRT Infrared Deep Sky Survey. The UKIDSS project is defined in \citet{law07}. UKIDSS uses the UKIRT Wide Field Camera \citep[WFCAM;][]{casa07}. The photometric system is described in \citet{hew06}, and the calibration is described in \citet{hodg09}. The pipeline processing and science archive are described in \citet{irwi04} and \citet{hamb08}. 
This publication makes use of data products from the Sloan Digital Sky Survey \citep{yor00}, the Two Micron All Sky Survey \citep{skr06}, the {\sl Wide-field Infrared Survey Explorer} \citep{wri10}, the VLT Survey Telescope ATLAS survey \citep{shan15}, and the Pan-STARRS1 survey \citep{cham16}. This publication has made use of data  from the VISTA Hemisphere Survey, ESO Progamme, 179.A-2010 (PI: McMahon) and the VIKING survey from VISTA at the ESO Paranal Observatory, programme ID 179.A-2004. Data processing has been contributed by the VISTA Data Flow System at CASU, Cambridge and WFAU, Edinburgh. 

Research has benefited from the M, L, and T dwarf compendium housed at DwarfArchives.org and maintained by Chris Gelino, Davy Kirkpatrick, and Adam Burgasser. This research has benefited from the SpeX Prism Spectral Libraries, maintained by Adam Burgasser at http://www.browndwarfs.org/spexprism. 

ZHZ and NL are partially funded by the Spanish Ministry of Economy and Competitiveness (MINECO) under the grants AYA2015-69350-C3-2-P. 
DH is supported by Sonderforschungsbereich SFB 881 `The Milky Way System' (subproject A4) of the German Research Foundation (DFG). FA received funding from the French `Programme National de Physique Stellaire' (PNPS) and the `Programme National de Plan\'etologie' of CNRS (INSU). The computations of atmosphere models were performed  in part on the Milky Way supercomputer, which is funded by the Deutsche Forschungsgemeinschaft (DFG) through the Collaborative Research Centre (SFB 881) `The Milky Way System' (subproject Z2) and hosted at the University of Heidelberg Computing Centre, and at the {\sl P\^ole Scientifique de Mod\'elisation Num\'erique} (PSMN) at the {\sl \'Ecole Normale Sup\'erieure} (ENS) in Lyon, and at the {\sl Gesellschaft f{\"u}r Wissenschaftliche Datenverarbeitung G{\"o}ttingen} in collaboration with the Institut f{\"u}r Astrophysik G{\"o}ttingen. 

\end{document}